\documentclass{article}
\usepackage{amssymb}
\usepackage{amsfonts}
\usepackage{amsmath}
\usepackage[doublespacing]{setspace}

\input{tcilatex}

\begin{document}

\title{A singular position-dependent mass particle in an infinite potential
well}
\author{Omar Mustafa$^{1}$ and S.Habib Mazharimousavi$^{2}$ \\
Department of Physics, Eastern Mediterranean University, \\
G Magusa, North Cyprus, Mersin 10,Turkey\\
$^{1}$E-mail: omar.mustafa@emu.edu.tr\\
\ Tel: +90 392 630 1314\\
\ \ Fax: +90 392 3651604\\
$^{2}$E-mail: habib.mazhari@emu.edu.tr}
\maketitle

\begin{abstract}
An unusual singular position-dependent-mass particle in an infinite
potential well is considered. The corresponding Hamiltonian is mapped
through a point-canonical-transformation and an explicit correspondence
between the target Hamiltonian and a P\"{o}schl-Teller type reference
Hamiltonian is obtained. New ordering ambiguity parametric setting are
suggested.


\end{abstract}

\section{Introduction}

In addition of being a descriptive model for some physical phenomena [1-36]
(including but not limited to, energy density many-body problem, electronic
properties of semiconductors), the position-dependent-mass (PDM) von Roos
Hamiltonian [33] (in $\hbar =2m_{\circ }=1$ units)%
\begin{equation}
H=-\frac{1}{2}\left[ m\left( x\right) ^{\alpha }\partial _{x}m\left(
x\right) ^{\beta }\partial _{x}m\left( x\right) ^{\gamma }+m\left( x\right)
^{\gamma }\partial _{x}m\left( x\right) ^{\beta }\partial _{x}m\left(
x\right) ^{\alpha }\right] +V\left( x\right)
\end{equation}%
is shown to be a mathematically useful model as it enriches the class of
exactly-solvable quantum mechanical problems. Whilst the parameters $\alpha
,\beta ,$ and $\gamma $ are subjected to the so-called von Roos constraint%
\begin{equation}
\alpha +\beta +\gamma =-1\text{ };\text{ \ }\alpha ,\beta ,\gamma \in 
\mathbb{R}
,
\end{equation}%
the only feasibly admissible case that ensures the continuity conditions at
the heterojunction boundaries between two crystals is for $\alpha =\gamma $.
This, in effect, reduces the domain of the acceptable values of the
so-called \emph{"ambiguity parameters"} $\alpha ,\beta ,$ and $\gamma $ and
suggests that the PDM-Hamiltonian (1) be rewritten as%
\begin{equation}
H=-m\left( x\right) ^{\alpha }\partial _{x}m\left( x\right) ^{\beta
}\partial _{x}m\left( x\right) ^{\alpha }+V\left( x\right) \text{ };\text{ \ 
}2\alpha +\beta =-1.
\end{equation}

Recently, moreover, Dutra and Almeida [14] have carried out a reliability
test on the orderings available in the literature. They have used an exactly
solvable Morse model and concluded that the orderings of Gora and Williams
[15] ($a=\beta =\gamma =0,$ $\alpha =-1$), and Ben Danial and Duke [16] ($%
a=\alpha =\gamma =0,$ $\beta =-1$) should be discarded for they result in
complex energies. Nevertheless, they have classified the ordering of Zhu and
Kroemer [17] ($a=0,$ $\alpha =\gamma =-1/2,$ $\beta =0$), and that of Li and
Kuhn [18] ($a=\alpha =0,\beta =\gamma =-1/2$) as good orderings. Therefore,
the continuity conditions at the heterojunction boundaries and Dutra's and
Almeida's [14] reliability test would ultimately single out Zhu and Kroemer (%
$a=0,$ $\alpha =\gamma =-1/2,$ $\beta =0$) as a \emph{"reliable good
ordering"} . Fitting into this category/classification, Mustafa and
Mazharimousavi [19] have used a PDM-pseudo-momentum operator and suggested a
new \emph{"reliable good ordering"} ($\beta =-1/2,$ $\alpha =\gamma =-1/4$).
Yet, in their study of classical and quantum PDM harmonic oscillator, Cruz
et al [20] have considered different ambiguity parameters settings and
argued that only one of these orderings, $\beta =-1/2,$ $\alpha =\gamma
=-1/4 $ (i.e., that of Mustafa and Mazharimousavi [19]), gives rise to a
potential term that is the same as the classical PDM oscillator. For more
details on the ambiguity associated with the uniqueness of the kinetic
energy operator, the reader my refer to [8, 14, 19, 20 and references cited
therein].

In this letter, only for mathematical and/or quantum mechanical curiosity on
the associated exact-solvability, we consider the PDM von Roos Hamiltonian
(1) for a quantum particle moving within the domain mandated by its own
PDM-function $m\left( x\right) $ accompanied by an infinite potential well $%
V\left( x\right) $. In section 2, we consider an unusual PDM-function of the
form [20]%
\begin{equation}
m\left( x\right) =\frac{m_{\circ }}{\left[ 1-\left( \frac{x}{a}\right) ^{2}%
\right] ^{2}}\text{ };\text{ }V\left( x\right) =\left\{ 
\begin{tabular}{l}
$0$ \ for $\left\vert x\right\vert <a$ \\ 
$\infty $ for $\left\vert x\right\vert \geq a$%
\end{tabular}%
\right. .
\end{equation}%
Under such settings, it is obvious that the classical motion of such a
particle is confined to the domain $\mathcal{D}\left( x\right) =\left(
-a,a\right) $. and the PDM-Hamiltonian in (1) would result a PDM-Schr\"{o}%
dinger equation of the form%
\begin{equation}
\left[ -\partial _{x}\frac{1}{m\left( x\right) }\partial _{x}+\tilde{V}%
\left( x\right) \right] \psi \left( x\right) =E\text{ }\psi \left( x\right) 
\text{ };\text{ \ }\left\vert x\right\vert <a,
\end{equation}%
with%
\begin{equation}
\tilde{V}\left( x\right) =\frac{1}{2}\left( 1+\beta \right) \frac{m^{\prime
\prime }\left( x\right) }{m\left( x\right) ^{2}}-\left[ \alpha \left( \alpha
+\beta +1\right) +\beta +1\right] \frac{m^{\prime }\left( x\right) ^{2}}{%
m\left( x\right) ^{3}}+V\left( x\right) .
\end{equation}%
Where primes denote derivatives with respect to $x$ and the PDM-Schr\"{o}%
dinger equation in (5) is known as the target equation. Obviously, as $%
\alpha ,\beta ,$ and $\gamma $ change within constraint (2), a profile
change of $\tilde{V}\left( x\right) $ in (6) is unavoidable (hence, an
ordering ambiguity conflict erupts in the process). Consequences of a
point-canonical-transformation (PCT) mapping (often mediates a transition
between two different effective potentials) on such a PDM particle are also
discussed in section 2. Therein, we shall witness that such a mass setting
results in an effective reflectionless modified P\"{o}schl-Teller type
potential well (cf. e.g., [20, 37]). We conclude in section 3.

\section{Consequences of PCT-mapping}

Following the well known PCT recipe (cf., e.g., [19,24;27]) would, with a
substitution of the form $\psi \left( x\right) =m\left( x\right)
^{1/4}\,\phi \left( q\left( x\right) \right) $ in (5), result in $q^{\prime
}\left( x\right) =\sqrt{m\left( x\right) }$ and suggests the following point
canonical transformation%
\begin{equation}
q\left( x\right) =\int\nolimits^{x}\sqrt{m\left( t\right) }dt=a\tanh
^{-1}\left( \frac{x}{a}\right) \text{ ; \ \ }\left\vert x\right\vert <a,
\end{equation}%
with $m\left( x\right) $ given in (4). Consequently, the PDM-Schr\"{o}dinger
equation (5) is mapped into%
\begin{equation}
\left( -\frac{d^{2}}{dq^{2}}+V_{eff}\left( q\right) \right) \phi \left(
q\right) =E\phi \left( q\right) \text{ ;\ \ \ }q\in \left( -\infty ,\infty
\right) ,
\end{equation}%
where%
\begin{equation}
V_{eff}\left( q\right) =g_{1}\frac{m^{\prime \prime }\left( x\right) }{%
m\left( x\right) ^{2}}-g_{2}\frac{m^{\prime }\left( x\right) ^{2}}{m\left(
x\right) ^{3}},
\end{equation}%
with 
\begin{equation}
\text{\ }g_{1}=\frac{1}{4}\left( 1+2\beta \right) \text{ },\text{ \ \ }%
g_{2}=\alpha \left( \alpha +\beta +1\right) +\beta +\frac{9}{16}.
\end{equation}%
One may then use (7) and substitute%
\begin{equation}
\frac{x}{a}=\tanh \left( \frac{q}{a}\right) \text{ ;\ \ \ }\left\vert
x\right\vert <a,
\end{equation}%
in (9) to imply%
\begin{equation}
V_{eff}\left( q\right) =-\frac{\lambda \left( \lambda -1\right) }{a^{2}\cosh
^{2}\left( q/a\right) }+\frac{8}{a^{2}}\left( 3g_{1}-2g_{2}\right) \text{ ; }%
\lambda \left( \lambda -1\right) =4\left( 5g_{1}-4g_{2}\right) .
\end{equation}%
Which, in turn, would lead to an over simplified reference Schr\"{o}dinger
equation of the form%
\begin{equation}
\frac{d^{2}\phi \left( q\right) }{dq^{2}}+\left( k^{2}+\mu ^{2}\frac{\lambda
\left( \lambda -1\right) }{\cosh ^{2}\left( \mu q\right) }\right) \phi
\left( q\right) =0\text{,}
\end{equation}%
where%
\begin{equation}
k^{2}=E-8\mu ^{2}\left( 3g_{1}-2g_{2}\right) =\mathcal{E}\text{ ; }\mu =1/a%
\text{\ .}
\end{equation}%
Remarkably, the "shifted-by-a-constant" potential well in (12) is known to
be reflectionless/transparent (at any energy) with a reflection coefficient
equals zero if $\lambda >1$ is a positive integer (cf., e.g., Lekner [21]
and/or Diaz et al. [25] for a comprehensive and detailed study on this
potential). Therefore, a quick recollection to the essentials of the
Modified P\"{o}schl-Teller potential by Diaz et al. [25] would (with a bound
state solution when $\mathcal{E}<0$ ) provide a clear correspondence to the
following recycled solution%
\begin{eqnarray}
\phi \left( q\right) &=&\left( \cosh \mu q\right) ^{\lambda }\left[
A\,_{2}F_{1}\left( a,b;\frac{1}{2};-\sinh ^{2}\mu q\right) +\right.  \notag
\\
&&\left. B\,\left( \sinh \mu q\right) \,_{2}F_{1}\left( a+\frac{1}{2},b+%
\frac{1}{2};\frac{3}{2};-\sinh ^{2}\mu q\right) \right]
\end{eqnarray}%
where 
\begin{equation}
a=\frac{1}{2}\left( \lambda -\frac{\sqrt{\left\vert \mathcal{E}\right\vert }%
}{\mu }\right) \,;\text{ \ \ \ }b=\frac{1}{2}\left( \lambda +\frac{\sqrt{%
\left\vert \mathcal{E}\right\vert }}{\mu }\right) ,
\end{equation}%
and%
\begin{equation}
\mathcal{E}_{n}\mathcal{=-}\mu ^{2}\left( \lambda -1-n\right) ^{2};\text{ \
\ }n\in 
\mathbb{N}
,\text{ }0\leq n<\lambda -1.
\end{equation}%
Hereby, as long as the reference Schr\"{o}dinger equation (13) is in point,
the corresponding solutions hold true. However, a mapping to the original
target problem defined through equations (4)-(6) would suggest that%
\begin{eqnarray}
\psi _{n}\left( x\right) &=&\left( \frac{1}{1-\left( \frac{x}{a}\right) ^{2}}%
\right) ^{\frac{\lambda +1}{2}}\left[ A\,_{2}F_{1}\left( a,b;\frac{1}{2};%
\frac{\left( \frac{x}{a}\right) ^{2}}{\left( \frac{x}{a}\right) ^{2}-1}%
\right) \right.  \notag \\
&&\left. +B\left( \sqrt{\frac{\left( \frac{x}{a}\right) ^{2}}{1-\left( \frac{%
x}{a}\right) ^{2}}}\right) \,_{2}F_{1}\left( a+\frac{1}{2},b+\frac{1}{2};%
\frac{3}{2};\frac{\left( \frac{x}{a}\right) ^{2}}{\left( \frac{x}{a}\right)
^{2}-1}\right) \right] ,
\end{eqnarray}%
and%
\begin{equation}
E_{n}=\mu ^{2}\left( 8\left( 3g_{1}-2g_{2}\right) -\left( \lambda
-1-n\right) ^{2}\right) ,\text{ \ \ }n\in 
\mathbb{N}
,\text{ }0\leq n<\lambda -1,
\end{equation}%
with%
\begin{equation}
a=\frac{1}{2}\left( 1+n\right) \text{ };\text{ \ \ }b=\lambda -\frac{1}{2}%
\left( 1+n\right) .
\end{equation}

The ground state solution of which would (with $n=0$) read%
\begin{eqnarray}
\psi _{0}\left( x\right) &=&A_{0}\left( \frac{1}{1-\left( \frac{x}{a}\right)
^{2}}\right) ^{\left( \lambda +1\right) /2}F\left( \frac{1}{2},\lambda -%
\frac{1}{2};\frac{1}{2};\frac{\left( \frac{x}{a}\right) ^{2}}{\left( \frac{x%
}{a}\right) ^{2}-1}\right)  \notag \\
&=&A_{0}\left( \frac{1}{1-\left( \frac{x}{a}\right) ^{2}}\right) ^{\left(
2-\lambda \right) /2},
\end{eqnarray}%
with an eigenvalue%
\begin{equation}
E_{0}=\mu ^{2}\left( 8\left( 3g_{1}-2g_{2}\right) -\left( \lambda -1\right)
^{2}\right) .
\end{equation}%
Nonetheless, the boundary condition on the wave function%
\begin{equation*}
\psi \left( x\right) =m\left( x\right) ^{1/4}\phi \left( q\left( x\right)
\right)
\end{equation*}%
on the $x$-axis would mandate%
\begin{equation}
\underset{x\rightarrow \pm a}{\lim }\psi \left( x\right) =0.
\end{equation}%
Feasibly, the satisfaction of this condition is obviously achievable when $%
\lambda >2$. This would imply that%
\begin{equation}
\lambda =\frac{1}{2}\left( 1+\sqrt{1+80g_{1}-64g_{2}}\right) >2.
\end{equation}%
At this point, we choose to stick with the continuity conditions at the
heterojunction boundaries between two crystals and take $\alpha =\gamma
\Rightarrow \beta =-1-2\alpha $ to imply%
\begin{equation}
\lambda =\frac{1}{2}\left( 1+\sqrt{\left( 3+8\alpha \right) ^{2}}\right)
>2\implies \left\vert 3+8\alpha \right\vert >3.
\end{equation}%
This suggests, in addition to von Roos constraint (2), new constraints $%
\alpha =\gamma <-3/4$ or $\alpha =\gamma >0$ on the ambiguity parameters. In
table 1, we list the parametric values associated with the available
orderings in the literature, along with their classifications as to being
admissible (i.e., $\lambda >2$ ) or non-admissible (i.e., $\lambda \leq 2$
). In the same table, we list some new orderings that are feasibly
admissible. \ Obviously, non of the orderings known in the literature can be
labeled as admissible, i.e., $\lambda >2$, (within our current methodical
proposal, of course).

\section{Concluding Remarks}

We have considered the PDM-von Roos Hamiltonian for a quantum particle
endowed with an unusual position-dependent mass function with two
singularities in an infinite potential well. A
point-canonical-transformation is used and an obvious correspondence between
two effective (reference and target) Hamiltonians is obtained. Apparently,
the singular PDM settings led to (through a PCT) the reflectionless modified
P\"{o}schl-Teller potential well. The exact solution of which is known in
different perspectives. We have followed Diaz et al. [25] to come out with
the result that non of the available known orderings in the literature is
labeled admissible. Hereby, it should be noted that, our classifications on
the "admissibility" and "non-admissibility" of the ordering ambiguity
parameters (reported in table 1) are consequences of the boundary condition
in (23). The validity of which resides only within the setting of the
current methodical proposal associated with our model in (4).

Nevertheless, one could re-scale $k^{2}$ of equation (2) of Lekner [21] for $%
\lambda =\nu +1$, for example, and deduce all related reflectionless
positive energy states, non-reflecting wave packets, etc. Yet, by re-scaling 
$k^{2}$ of equation (39.2) of Fl\"{u}gge [37] for $\lambda =\nu +1$, one
would recycle Fl\"{u}gge's results. Consequently, we may conclude that the
ordering ambiguity conflict associated with the uniqueness of the kinetic
energy operator does not only depend on the heterojunction boundaries
between two crystals and the Dutra's and Almeida's [14] reliability test.
Even the form of the potential and/or the form of the position dependent
mass may have their say in the process, so to speak. Such ambiguity in the
non-uniqueness representation of the position-dependent mass Schr\"{o}dinger
Hamiltonian should be attributed to the lack of the Galilean invariance
(cf., e.g., ref.[33] for more details on the issue).

To the best of our knowledge, such a study has not been considered
elsewhere, not only within the Hermitian PDM-Hamiltonians' settings, but
neither within the complex non-Hermitian settings. A gap that remains
interesting and merits exploration/attention.

.

\newpage

\begin{table}[tbp]
\caption{The parametric values associated with the available orderings in
the literature, along with their classifications as to being admissible
(i.e., $\protect\lambda >2$ ) or non-admissible (i.e., $\protect\lambda \leq
2$ ) . New admissible values are also suggested.}
\begin{center}
\vspace{1cm}%
\begin{tabular}[t]{|c|c|c|c|c|c|}
\hline
Available ordering & $\alpha =\gamma $ & $\beta $ & $\lambda $ & $E$ & 
Admissibility \\ \hline
\begin{tabular}{l}
Zhu and Kroemer [17], \\ 
and Li and Kuhn [18]%
\end{tabular}
& $-\frac{1}{2}$ & $0$ & $1$ & $\frac{1}{4}$ & non-admissible \\ \hline
Gora and Williams [15] & $-$ & $0$ & $-$ & $-$ & non-admissible \\ \hline
Ben Danial and Duke [16] & $0$ & $-1$ & $2$ & $0$ & non-admissible \\ \hline
Mustafa and Mazharimousavi [19] & $-\frac{1}{4}$ & $-\frac{1}{2}$ & $1$ & $0$
& non-admissible \\ \hline
New & $-1$ & $1$ & $3$ & $5$ & admissible \\ \hline
New & $\frac{1}{4}$ & $-\frac{3}{2}$ & $3$ & $0$ & admissible \\ \hline
New & $\frac{1}{2}$ & $-2$ & $4$ & $0$ & admissible \\ \hline
New & $\frac{3}{4}$ & $-\frac{5}{2}$ & $5$ & $0$ & admissible \\ \hline
New & $1$ & $-2$ & $6$ & $0$ & admissible \\ \hline
\end{tabular}%
\end{center}
\end{table}

\end{document}